# Minimum Magnetizability Principle

Akhilesh Tanwar[1], Debesh Ranjan Roy[2], Sourav Pal[1,*] and Pratim Kumar Chattaraj[2,*]

*[1]Theoretical Chemistry Group, Physical Chemistry Division, National Chemical Laboratory, Pune-411008, India*

*[2]Department of Chemistry, Indian Institute of Technology, Kharagpur 721302, India*

 E-mail: s.pal@ncl.res.in, pkc@chem.iitkgp.ernet.in

In this communication we propose and also verify through *ab initio* calculations a new electronic structure principle, viz. the minimum magnetizability principle (MMP), to extend the domain of applicability of the conceptual density functional theory (DFT) [1] in explaining the magnetic interactions and magnetochemistry. This principle may be stated as, "A stable configuration/conformation of a molecule or a favorable chemical process is associated with a minimum value of the magnetizability". We also show that a soft molecule is easily polarizable and magnetizable than a hard one.

Chemical reactivity parameters such as electronegativity[2] (χ) and hardness[3,4] (η) have been defined within DFT[1] as follows:

$$\chi = -\mu = -\left(\frac{\partial E}{\partial N}\right)_{v(\vec{r})} \qquad (1)$$

and $$\eta = \frac{1}{2}\left(\frac{\partial^2 E}{\partial N^2}\right)_{v(\vec{r})} = \frac{1}{2}\left(\frac{\partial \mu}{\partial N}\right)_{v(\vec{r})} \qquad (2)$$

where, μ is the chemical potential of an N- electron system with external potential *v*(r) and total energy E. A many-particle system is completely characterized by N and *v*(r). While χ and η describe the response of the system when N changes (ΔN perturbation) at fixed *v*(r), the linear density response function does that job for the variation of *v*(r) (Δ*v* perturbation) at constant N. The linear responses of the electronic cloud of a chemical species to weak external electric and magnetic fields are measured in terms of polarizability (α) and magnetizability (ξ) respectively, as follows:

$$P = \alpha \in \quad ; \quad \alpha = -\frac{\partial^2 E(\in)}{\partial \in^2}\Big|_{\in=0} \qquad (3)$$

and $$m = \xi B \quad ; \quad \xi = -\frac{\partial^2 E(B)}{\partial B^2}\Big|_{B=0} \qquad (4)$$

where P and m refer to induced dipole moment and magnetic moment respectively. ∈ and B refer to the external electric and magnetic fields respectively. Magnetizability of the system can be decomposed into its diamagnetic component ($\xi_{dm}$) and paramagnetic component ($\xi_{pm}$). $\xi_{dm}$ is negative.

$$\xi_{Total} = \xi_{dm} + \xi_{pm} \qquad (5)$$

Two hardness related electronic structure principles are the hard-soft acid-base (HSAB) principle[5] and the maximum hardness principle (MHP)[6] while the former states that,[5] "Hard likes hard and soft likes soft", the statement of the latter is,[6] "There seems to be a rule of nature that molecules arrange themselves so as to be as hard as possible". Parr et al defined[7] electrophilicity index as follows:

$$\omega = \mu^2 / 2\eta \qquad (6)$$

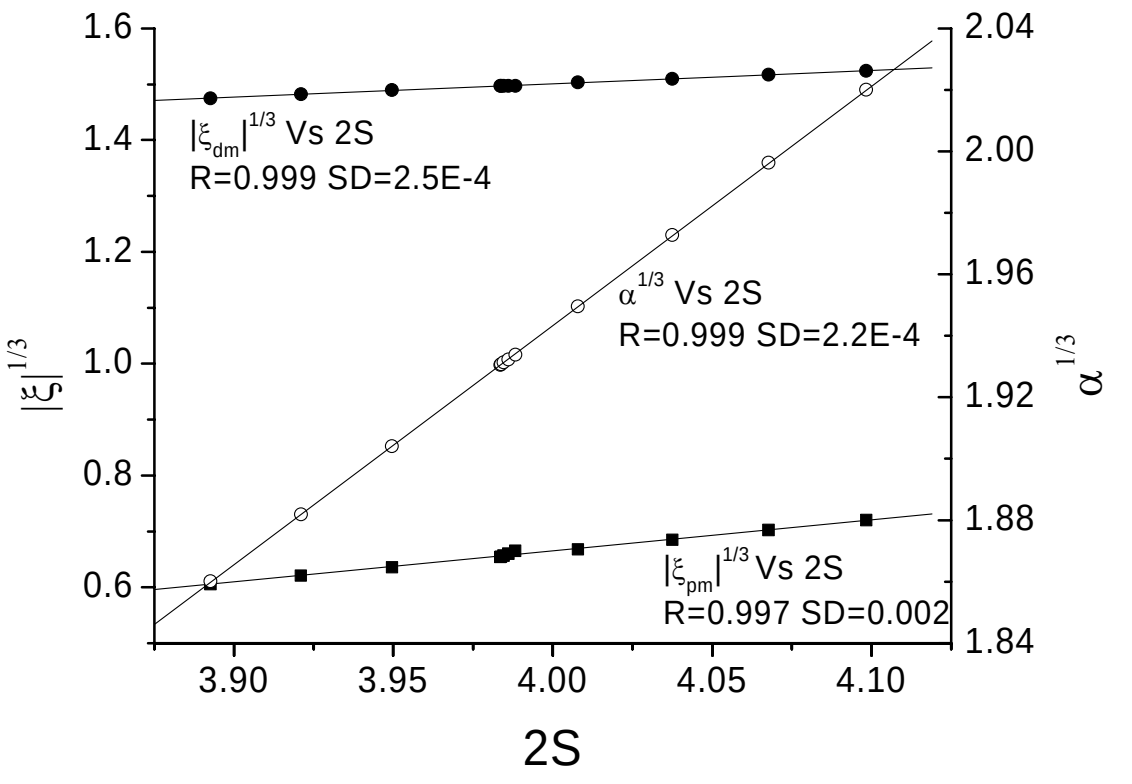

***Figure 1:*** Plots of $|\xi|^{1/3}$ and $\alpha^{1/3}$ vs 2S calculated during the stretching of water [MP2/6-31++G**].

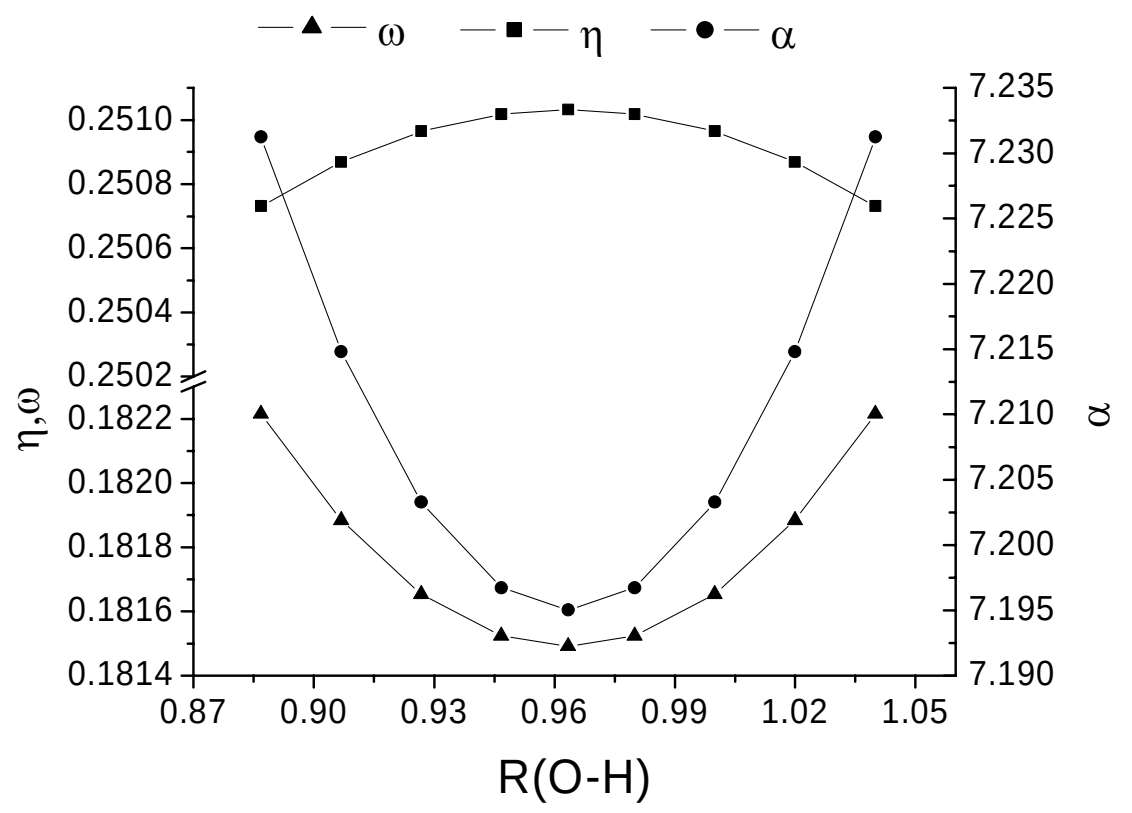

***Figure 2:*** Profiles of η, α, and ω during the asymmetric stretching of water [MP2/6-31++G**]. η,ω,α in a.u. and R(O-H) in Å

For a stable configuration with proper extremal behavior of both μ and η, the electrophilicity[7] also assumes its minimum value.[8] The global softness is defined as[1]

$$S = 1 / 2\eta \qquad (7)$$

There exists a linear relationship between softness and $\alpha^{1/3}$ for both ground[9] and excited[10] electronic states. This linear relationship gave rise to the minimum polarizability principle (MPP)[11] which states that, "The natural direction of evolution of any system is towards a state of minimum polarizability". In this communication we propose a minimum magnetizability principle as, "A stable configuration/conformation of a molecule or a favorable chemical process is associated with a minimum value of the magnetizability".

In order to verify our prognosis we study the asymmetric and symmetric stretching of $H_2O$. Calculations were performed using DALTON[12] system of programs. MP2 level properties are calculated using second-order-perturbation-propagation-approximation (SOPPA) [13] method.

Figure 1 depicts the linear variations in $\alpha^{1/3}$, $|\xi_{dm}|^{1/3}$ and $|\xi_{pm}|^{1/3}$ with 2S during the stretching of $H_2O$. The variants of HSAB principle and MHP in terms of polarizability and magnetizability would be helpful in analyzing chemical reactions especially when they are associated with changes in electrical and magnetic properties.

Figure 2 presents the related variations in η, α and ω with respect to (O-H) bond length during asymmetric stretch. As expected from the MHP and MPP, η attains its maximum value and α attains its minimum value for the equilibrium bond length where ω is also a minimum. The chemical potential remains constant[14] in the sense of Pearson-Palke[15] and Makov.[16] During symmetric stretch no maximum in η is observed near equilibrium geometry as chemical potential itself is not constant hence the principle cannot be tested in this case. This is also in agreement with earlier works on PMH.[14, 15] Variation in μ, η, α, ξ and ω with respect to (O-H) bond length during symmetric stretch is available as supporting information.

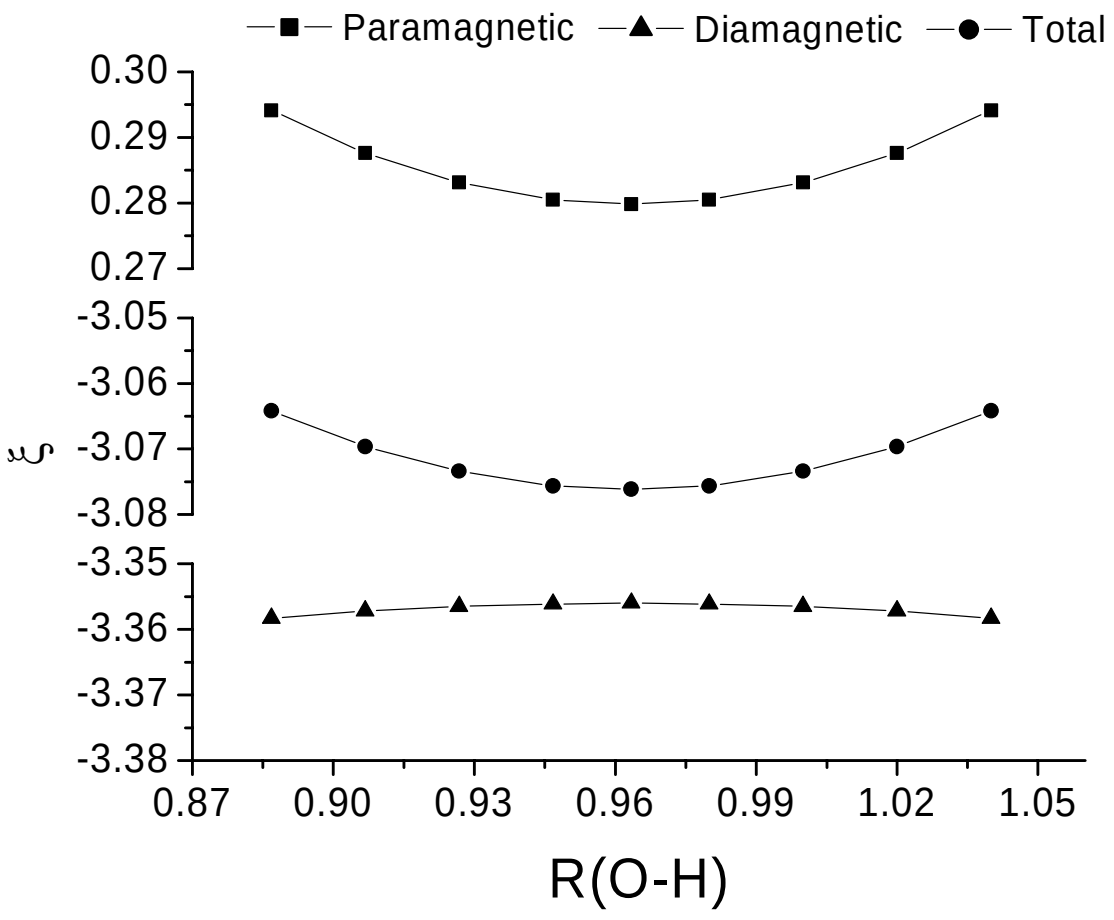

***Figure 3:*** Plot of ξ during asymmetric stretching of water[MP2/6-31++G**]. ξ in a.u. and R(O-H) in Å

Figure 3 presents the variation in magnetizabilities of water during asymmetric stretch. It is heartening to note that in concurrence with MMP stated earlier, total magnetizability indeed shows its minimum at equilibrium geometry. A minimum in the total magnetizability is driven by the minimum observed in the paramagnetic component. The diamagnetic component, however, does not change much during the asymmetric stretch.

This principle (MMP) will help understanding the magnetic interactions better.

**Acknowledgment.** Financial assistance from CSIR, New Delhi is gratefully acknowledged.

**Supporting Information available:** Variation in μ, η, α, ξ and ω with respect to (O-H) bond length during symmetric stretch. This material is available free of charge via the internet at http://pubs.acs.org.